\documentclass[12pt]{article}
\usepackage{amssymb}
\usepackage[dvips]{epsfig}
\usepackage[utf8]{inputenc}

\newcommand{\be}{\begin{equation}}
\newcommand{\ee}{\end{equation}}

\newcommand{\bn}{\begin{eqnarray}}
\newcommand{\en}{\end{eqnarray}}

\def\bea{\begin{eqnarray}}
\def\eea{\end{eqnarray}}

\newcommand{\beq}{\begin{eqnarray}}
\newcommand{\eeq}{\end{eqnarray}}

\usepackage[utf8]{inputenc}
\usepackage[dvips]{geometry}

\usepackage{indentfirst}
\usepackage{amsmath}
\usepackage{graphicx}
\usepackage{amssymb}
\def\({\left(}
\def\){\right)}

\def \= {\,\dot{=}\,}
\begin{document}

\title{\textbf{Dynamical mass generation in QED$_3$: \\ A non-perturbative approach}}
\author{ 
G. B. de Gracia\footnote{gb9950@gmail.com},\ \quad B. M. Pimentel \footnote{b.m.pimentel@gmail.com},\ \quad L. Rabanal \footnote{luis.rabanal@unesp.br} \\ 
\textit{{Instituto de F\'{i}sica Te\'{o}rica (IFT), Universidade Estadual Paulista (UNESP)} }\\
\textit{{Rua Dr. Bento Teobaldo Ferraz, 271,
Bloco II, Barra Funda} }\\
\textit{{CEP 01140-070-S\~{a}o Paulo, SP, Brazil} } \\}
\date{}
\maketitle

\begin{abstract}
In this work we provide a non-perturbative description of the phenomenon of dynamical mass generation in the case of quantum electrodynamics in $2+1$ dimensions. We will use the Kugo-Ojima-Nakanishi formalism to conclude that the physical Hilbert space of the asymptotic photon field is the same as that of the Maxwell-Chern-Simons.
\end{abstract}

\section{Introduction}
It is widely known that quantum electrodynamics in $2+1$ dimensions (QED$_3$) has important applications in condensed matter physics. The predominant example is the quantum Hall effect (QHE), where a pure topological Chern-Simons (CS) term \cite{hall} is commonly added to model the response of the quantum Hall ground state to low energy perturbations as an effective theory~\cite{hall1}, but it can also be used to study the behavior of ultracold matter in optical lattices~\cite{cond}. Nevertheless, this theory also has outstanding properties from the theoretical point of view.

What is special about $2+1$ spacetime dimensions? Let us consider first the theory in the absence of fermions. By naive dimensional analysis we note a big difference with respect to the $3+1$ case. The vector potential $A_{\mu}$ has dimension 1 (in units of mass) in any $d$-dimensional spacetime. As a consequence, if we write the Lagrangian in the form
\begin{equation}
	\mathcal{L}_{QED_d} = -\frac{1}{4e^2}F_{\mu\nu}F^{\mu\nu} + A_{\mu}J^{\mu},
	\label{qedaction}
\end{equation}
then we realize that the coupling constant $e^2$ is dimensionless in $3+1$ but dimensionful in other dimension $d \ne 3+1$. In particular, in $d = 2+1$, the effective dimensionless coupling would be $e'^2 = e^2/E$, where $E$ is the energy scale. In the ultraviolet (UV) regime, $E$ tends to infinity and the coupling $e'$ goes to zero implying that the theory is superrenormalizable and  always asymptotically free, i.e., this theory describes \textit{free photons in the UV}. In this sense, the UV does not matter at all. On the other hand, the theory is always \textit{strongly coupled in the infrared} (IR) because $e' \rightarrow \infty$ as $E\rightarrow 0$. Consequently, the IR limit of the theory becomes a playground for developing ideas to tackle more realistic problems as confinement in quantum chromodynamics (QCD) \cite{herbut,grignani1,grignani2} or gapped boundary phases in topological insulators (TI) \cite{seibergwitten}.

Another interesting property of the theory in this dimensionality is related to the existence of magnetic monopoles. Whenever we have a $U(1)$ gauge field we have a new current

\begin{equation}
	\mathcal{J}^{\mu} \propto \epsilon^{\mu\nu\rho}F_{\nu\rho},
	\label{TopCurr}
\end{equation}
which is identically conserved without imposing the equations of motion, i.e., it is not a Noether current. Its conservation is equivalent to the Bianchi identity $dF = 0$, where $F$ is the two-form field strength. This follows simply by the symmetry of partial derivatives which contributes to zero when contracted with a Levi-Civita symbol if $A_\mu$ is globally well-defined. A natural question is: Who is charged under the charge
\begin{equation}
	\mathcal{Q} = \int d^2x \mathcal{J}^0?
	\label{TopCharg}
\end{equation}
If we replace (\ref{TopCurr}) in (\ref{TopCharg}) we obtain that $\mathcal{Q}$ is equal to a magnetic flux from which we conclude that \textit{magnetic monopoles are charged under} $\mathcal{Q}$. This charge is known as the vortex charge because Abrikosov-Nielsen-Olesen (ANO) vortices carry it when the theory is put in the Higgs phase \cite{borokhov}. 
Moreover, \textit{the vector potential} $A_{\mu}$ \textit{can be dualized to a free scalar} $\sigma$ \textit{in the UV}. It is known as the dual photon field. The construction of the dual theory is carried out analogously as the electric-magnetic duality of Maxwell theory in $3+1$ spacetime dimensions, namely,
\begin{equation}
	Z = \int \mathcal{D}A_{\mu} \exp\left(-\int_x\frac{F^2}{4e^2}\right) \rightarrow \int \mathcal{D}\sigma\mathcal{D}F_{\mu\nu} \exp\left[\int_x \left(-\frac{F^2}{4e^2} + \frac{i}{4\pi} \sigma \epsilon^{\mu\nu\rho}\partial_{\mu}F_{\nu\rho}\right)\right],
\end{equation}
where the dual photon $\sigma$ has been introduced as a Lagrange constraint in order to be able to treat the field strength as the integration variable \cite{polchinski}. After integrating out the field strength through its equation of motion we obtain
\begin{equation}
	Z_{\text{dual}} = \int D\sigma \exp\left( -\int_x \frac{e^2}{8\pi^2}(\partial\sigma)^2 \right).
	\label{DualPhotonPathIntegral}
\end{equation}
It can be shown straightforwardly that the conserved Noether current of this dual theory under the shift symmetry $\sigma \rightarrow \sigma + \text{const}$, coincides with the current (\ref{TopCurr}). This in turn implies that $F_{\alpha\beta} \propto \epsilon_{\alpha\beta\mu}\partial^{\mu}\sigma$. Consequently, $\partial^{\alpha}F_{\alpha\beta} = 0$ and the theory describes free photons in accordance with our previous discussion of the UV.

We can also add to the action (\ref{qedaction}) Chern-Simons (CS) or topological terms. Although, these terms do not described any dynamics and have zero degrees of freedom, they can have effects on the degeneracy of the ground state of the theory with interesting consequences \cite{chen}. When added, the theory is known as Maxwell-Chern-Simons (MCS) theory and it is gapped, i.e., \textit{the photon is massive}. After having understood that the theory is strongly coupled in the IR, we could, effectively, drop out the Maxwell term and conclude that \textit{the theory is a topological quantum field theory} (TQFT) \textit{in the IR limit} \cite{dunne}.

Now, interesting things start happening when matter (either fermions, bosons or both) is taken into consideration. In the above-mentioned effective description of the theory in the IR, we can write
\begin{equation}
	\mathcal{L} = \mathcal{L}_{\text{CS}} + \mathcal{L}_{\text{Fermion}} \quad \text{or} \quad 	\mathcal{L} = \mathcal{L}_{\text{CS}} + \mathcal{L}_{\text{Scalar}},
\end{equation}
because the Maxwell term disappears. Obviously, \textit{all dynamics arise from matter}. However, they are no longer TQFT but \textit{believed to be}\footnote{In fact, computations of the IR properties of the theory using Schwinger-Dyson equations show that they might not be non-trivial CFTs \cite{pisarski}. However, it is common to argue that these kind of conclusions are based on truncation methods. A similar argument could be pointed out against the functional renormalization group (FRG) method \cite{gies}.} non-trivial conformal field theories (CFT) when their masses are tune to zero in a IR fixed point. If this were true, there is a possibility of studying topological changing phase transitions by relevant deformations, e.g., mass deformations, between TQFT's,
$$TQFT_1 \xleftarrow{\text{Relev. Deform.}} CFT \xrightarrow{\text{Relev. Deform.}} TQFT_2.$$
Yet, there is no free lunch. There is a subtlety with massive fermions in $2+1$ dimensions. Their path integral description presents \textit{parity anomaly}, that is, parity is a symmetry at the classical level but is not at quantum level \cite{witten}. Among the several ways this anomaly can arise, one can understand it through the $1$-loop term in the low energy approximation of the Euclidean path integration \cite{alvarez,niemi,redlich}
\begin{equation}
	-\frac{1}{2} \text{Tr} \left( \frac{1}{i\gamma_{\mu}\partial_{\mu} - m_e}\gamma_{\nu}A_{\nu} \frac{1}{i\gamma_{\mu}\partial_{\mu} - m_e}\gamma_{\delta}A_{\delta}\right) = \frac{1}{2}\int \frac{d^3p}{(2\pi)^3} A_{\mu}(p)\Gamma_{\mu\nu}(p,m_e)A_{\nu}(p),
\end{equation}
with
\begin{equation}
	\Gamma_{\mu\nu}(p,m_e) = -\int \frac{d^3k}{(2\pi)^3} \frac{\text{Tr}\left[\gamma_{\mu}(\gamma_{\rho}(p_{\rho}+k_{\rho})+m_e)\gamma_{\nu}(-\gamma_{\delta}k_{\delta}-m_e)\right]}{\left[(p+k)^2+m_e^2\right]^2(k^2+m_e^2)^2}, \quad p \ll m_e.
\end{equation}
At zero temperature, the contribution of the anomaly to the effective action (in Minkowski signature) is of the form
\begin{equation}
	S_{\text{eff}}[A,m_e]^{(T=0)} = \cdots + \frac{1}{2} \frac{1}{4\pi} \frac{m_e}{|m_e|} \int d^3x \  \epsilon^{\mu \nu \beta}A_\mu \partial_\nu A_\beta + \cdots,
\end{equation}
whereas at finite temperature, after imposing the anti-periodic conditions of Dirac fermions $\psi(0,\boldsymbol{x}) = - \psi(\beta,\boldsymbol{x})$, we obtain
\begin{equation}
	S_{\text{eff}}[A,m_e]^{(T\ne0)} = \cdots + \frac{1}{2} \frac{1}{4\pi} \frac{m_e}{|m_e|} \tanh\left(\frac{|m_e|}{T}\right)\int_0^{1/T} dt\int d^2x \ \epsilon^{\mu\nu\rho}A_{\mu}\partial_{\nu}A_{\rho} + \cdots.
\end{equation}
Clearly, CS terms have arisen and the breaking of parity depends on the sign of $m$.

A detailed study of all the above-mentioned points and additional exact results in lattice models, e.g, weak duality \cite{kramers,peskin}, have led to the conjeture of the existence of a web of dualities in $2+1$ spacetime dimensions with possible connections to the realization of 3D bosonization \cite{webduality1,webduality2}. Hence, this theory deserves to be investigated further.

This work investigates some properties of QED$_3$ within the covariant operator formalism of quantum field theory. We call it the Kugo-Ojima-Nakanishi (KON) formalism \cite{kugoojima,Nakanishi,Nak1,Laut}. Firstly, we want to address the problem of a dynamical mass generation for the photon arising from the interaction with generic charged particles, that is, either bosons or fermions in a given specific representation. In fact, this phenomenon is expected to happen since in this dimension the appearance of a mass term is in accordance with the local symmetries of the theory if one considers a discrete symmetry breaking scenario, e.g., parity anomaly in the presence of massive fermions.

The standard way to gap the photon is considering the MCS theory from the outset. In other words, by ``adding by hand'' a bare topological mass term. However, we argue that this procedure is not necessary and that QED$_3$ per se provides us these terms \textit{dynamically}. Under this perspective the conventional low energy quantum Hall effect field description \cite{mar, tong} would arise naturally from the situation of bidimensional electrons interacting with initially massless photons. The interaction changes the dispersion relation of the photon and the electromagnetic correlations become to fall faster through the material medium.

This property of QED$_3$ was intensively studied within perturbation theory (PT). In this framework, however, it was uncertain whether a renormalized mass of the photon actually existed. If the Pauli-Villars regularization method was used, the photon could either acquire an effective mass or remain masless which, by themselves, are two contradictory results. In fact, this kind of problem appears in the conventional perturbation theory when the regularization techniques are wrongly applied. Nevertheless, in \cite{Pim2} it was shown that if Pauli-Villars regularization is correctly applied, no problem arises and the photon becomes massive. The controversy was finally completely solved (of course, only in PT) by using the causal perturbation theory \cite{Pim1} where by construction no regularization is needed.

This paper is organized as follows. In section 2 we consider, following the ideas of \cite{Nakanishi}, the ``pre-Maxwell-Chern-Simons model'' to derive the non-perturbative two-point function of the gauge field, a ``massive'' combination of field operators, an asymtoptic constraint for the matter currents, and a general condition for the existence of a renormalized mass of the photon with arbitrary matter currents. In section 3 we compare our result with the one obtained from PT for the particular case of fermionic matter in bidimensional representation. Finally, in section 4, the asymtoptic structure is constructed revealing that our ``massive'' combination has indeed a dynamically generated massive character. The conclusions and the outlook are presented in section 5. The metric signature $+--$ is used throughout.

\section{Effective mass of the photon in 2+1 dimensions}
Let us start with the following Lagrangian density within the KON formalism
\begin{equation}
	\mathcal{L} = -\frac{1}{4}F_{\mu \nu}F^{\mu \nu} + \frac{m}{4}\epsilon^{\mu \nu \rho}F_{\mu\nu}A_{\rho} + B\partial^\mu A_\mu + \frac{1}{2}\alpha B^2 + J^{\mu}A_{\mu} + \mathcal{L}_M.
	\label{eq:LagrangianGeneralCurrentPMCS}
\end{equation}
In the above expression, $\mathcal{L}_M$ is a generic matter Lagrangian density, $J^{\mu}$ is an \textit{arbitrary} $2+1$ dimensional matter current that breaks parity and $B$ is an auxiliary field that keeps track of the gauge fixing condition via the gauge parameter $\alpha$. Needless to say, $\mathcal{L}$ is invariant under the gauge transformations
\begin{equation}
	\delta A_\mu(x) = \partial_\mu \Lambda(x), \qquad \Box \Lambda = 0, \qquad \delta B(x) = 0,
\end{equation}
wherein $\Lambda$ is a c-number. 

We are interested in the behavior of this theory in the limit $m\rightarrow 0$. In 3 + 1 spacetime dimensions without spontaneous symmetry breaking (SSB), the renormalized mass of the photon is constrained to vanish as the bare mass goes to zero. This follows by the Johnson's theorem \cite{jon}. We want to follow this line of thought in 2 + 1 dimensions in order to show that in the limit $m\rightarrow 0$ a renormalized mass $m_r$ for the photon exists. It is the aim of this paper to derive a general mathematical expression for this statement (cf. (\ref{eq:MassRenormalization})). 

The Heisenberg equations of motion read
\begin{align}
	\partial^{\mu}A_{\mu} + \alpha B &= 0 \label{eq:GaugeCondition}\\
	\partial_{\mu} F^{\mu\nu} + m\epsilon^{\nu\mu\beta}\partial_{\mu}A_{\beta} - \partial^{\nu} B &= -J^{\nu}  \label{eq:EquationForCurrent}\\
	\partial_{\mu}J^{\mu} = 0. \label{eq:CurrentConserv}
\end{align}
Applying $\partial_{\nu}$ to (\ref{eq:EquationForCurrent}) and using (\ref{eq:CurrentConserv})
we determine the equation of motion for the $B$-field
\begin{equation}
	\Box B = 0.
	\label{eq:B-Field}
\end{equation}
Hence, as usual in the case of an Abelian theory, the subsidiary condition necessary to identify the physical space $\mathfrak{F}_{\text{phys}}$ is given by
\begin{equation}
	B^+(x) | \text{phys} \rangle = 0, \quad \forall | \text{phys} \rangle \in \mathfrak{F}_{\text{phys}}.
\end{equation}

In order to give a non-perturbative description of the dynamical mass generation phenomenon, let us first determine the vacuum expectation values of the commutation relations of the Heisenberg fields $A_{\mu}$. Equal-time commutation relations, quantum equations of motion and symmetries is all what we need. Although, an exact answer for them in the presence of interactions is almost impossible, the spectral representation method helps us to extract valuable information. In particular, it guides the construction of the asymptotic fields of the theory which represent the in/out Fock spaces $\mathfrak{F}$\footnote{We will write $\mathfrak{F}$ for both spaces in the assumption of asymptotic completeness, i.e., no bound states will emerge in the asymptotic region.}.

Since the matter current is gauge invariant it has vanishing projection with the auxiliary $B$-field, that is, $\left[ J^\mu(x), B(y) \right]=0$ or $J^{\mu}(x)|0\rangle \in \mathfrak{F}_{\text{phys}}$. From this, together with the sourced equations of motion and the zero norm character of $B(x)$, we find that (see Appendix)
\begin{equation}
\bigg( \Box^x\eta^{\alpha\nu} + m\epsilon^{\alpha\mu\nu}\partial_{\mu}^x \bigg) \bigg( \Box^y\eta^{\beta\sigma} + m\epsilon^{\beta\mu\sigma}\partial_{\mu}^y \bigg) \big[ A_{\nu}(x), A_{\sigma}(y) \big] = \left[ J^{\alpha}(x), J^{\beta}(y) \right].
\label{RelationBetweenSpectralFunctions}
\end{equation}
This result means that the spectral function for the full two-point function of the gauge field are related to the corresponding spectral function of the arbitrary matter current. In particular, the asymptotic structure of the latter imposes constraints on the former. A useful constraint can be derived by applying a trick based on reference \cite{des}.
Considering a renormalized mass $m_r$, we can find an asymtoptic parity breaking condition for the current and a pure massive physical discrete pole excitation by means of the expression
\begin{equation}
	\Big(\Box+m^2_r \Big)\mathcal{U}^{\mu}=\Big( m_rJ^{\mu}+\epsilon^{\mu \alpha \nu}\partial_\alpha J_\nu \Big).
	\label{Trick}
\end{equation}
If the asymptotic field $\mathcal{U}_{\mu}$ is to describe a purely massive field then it must satisfy the Proca conditions
\begin{equation}
	\quad \partial_\mu \  \mathcal{U}^{\mu}=0 \quad \text{and} \quad     \left(\Box+m^2_r\right)\mathcal{U}^{\mu}=0,
\end{equation}
and must be physical in the following sense
\begin{equation}
	\left[ {\cal{U}}^\mu(x), B(y) \right] = 0.
\end{equation} 
Hence, an asymptotic condition for the matter current follows immediately
\begin{equation}
	\epsilon_{\mu \nu \alpha}\partial^\nu J^{\alpha}_{ \text{as}}=-m_r J_\mu^{\text{as}}.
	\label{AsymptoticConditionForCurrents}
\end{equation}
This constraint will help us to fix some constants below whereas the identification of the asymptotic field is devoted to section 4.

Going back to equation \eqref{RelationBetweenSpectralFunctions}, we can find a general result for the vacuum expectation value of the gauge field commutator as follows
\begin{align}
	\langle 0 | \left[ A_{\mu}(x), A_{\nu}(y) \right] | 0 \rangle &= a\left(\eta_{\mu\nu} + \frac{1}{m^2}\partial_{\mu}\partial_{\nu} - \frac{1}{m}\epsilon_{\mu\nu\sigma}\partial^{\sigma} \right)\Delta(x-y;m^2) \nonumber \\
	&\quad+ \left( b\partial_{\mu}\partial_{\nu} + c\epsilon_{\mu\nu\beta}\partial^{\beta} \right) \Delta(x-y;0) + f\partial_{\mu}\partial_{\nu}E(x-y;0) 	 \nonumber \\
	&\quad- i\int^{\infty}_{0}ds \left[ \rho(s) \left( \eta_{\mu\nu} + s^{-1}\partial_{\mu}\partial_{\nu} \right) + \widetilde{\rho}(s) \epsilon_{\mu\nu\beta}\partial^{\beta} \right] \Delta(x-y;s),
	\label{eq:SpectralRepresentation1}
\end{align}
where the Green's functions are defined by the following Cauchy data
\begin{align} 
	\Box \Delta(x-y; s) &= -s\Delta(x-y; s), \quad \Delta(x-y; s)|_0 = 0, \quad \partial_0^x\Delta(x-y; s)|_0 = -\delta^2(x-y) \\
	\big(\Box+s\big) E(x-y; s) &= \Delta(x-y; s), \quad E(x-y; s)|_0=0, \quad (\partial_0^x)^3E(x-y;s)|_0 = -\delta^2(x-y),
\label{CauchyData}
\end{align}
with the subscript $|_0$ meaning $|_{x_0=y_0}$. In fact, the first two lines in \eqref{eq:SpectralRepresentation1} belong to the kernel of the differential operator in the left-hand side of \eqref{RelationBetweenSpectralFunctions}, that is, it is the solution in the absence of matter currents. The last term is the non-homogeneous part of the solution which arises due to the presence of matter currents, its specific form is fixed by current conservation (\ref{eq:CurrentConserv}).

By imposing the gauge fixing condition (\ref{eq:GaugeCondition}), the relation $f=-i\alpha$ is obtained. Using the initial condition $[A_k(x),\partial_0 A_l(y)]|_{0}=-i\eta_{kl}\ \delta^2(x-y)$ we have
\begin{equation}
	-i = -a - i\int^\infty_{0+}ds \ \rho(s), \quad\quad \frac{a}{m^2} + b = i\int^\infty_{0}ds \ s^{-1}\rho(s),
	\label{spectral1}
\end{equation}
and using $[A_k(x),A_l(y)]|_{0} = 0$ we have
\begin{equation}
    c - \frac{a}{m} = i\int^\infty_{0}ds \ \tilde{\rho}(s).
\end{equation}
These results have been completely general so far but we can study particular solutions of them motivated by physical facts. Henceforth, we shall fix $a=0$, as is done in the spontaneous symmetry breaking context \cite{Nak2}, since in the MCS theory as well as in QED$_3$ there is just one asymptotic transverse physical excitation with a given mass. If $a\ne 0$, it would imply the existence of an additional asymptotic particle in the physical sector besides the radiatively generated one, namely, the one when parity breaking matter fields are considered in consistency with the Wilsonian perspective. However, this conclusion leads to a violation of the number of degrees of freedom in the theory and, thus, it is not allowed. Consequently,
\begin{equation}
	b = i\int^\infty_{0}ds \ s^{-1}\rho(s), \qquad c = i\int^\infty_{0}ds \ \tilde{\rho}(s), \qquad \int^{\infty}_{0^+} ds \ \rho(s) = 1.
	\label{spectral2}
\end{equation}

All in all, we obtain the following non-perturbative result
\begin{align}
	\langle 0 | \left[ A_{\mu}(x), A_{\nu}(y) \right] | 0 \rangle &= i \left( L\partial_{\mu}\partial_{\nu} + R\epsilon_{\mu\nu\beta}\partial^{\beta} \right) \Delta(x-y;0) -i\alpha\partial_{\mu}\partial_{\nu}E(x-y;0) 	\nonumber \\
	&\quad- i\int^{\infty}_{0}ds \left[ \rho(s) \left( \eta_{\mu\nu} + s^{-1}\partial_{\mu}\partial_{\nu} \right) + \widetilde{\rho}(s) \epsilon_{\mu\nu\beta}\partial^{\beta} \right] \Delta(x-y;s)
	\label{eq:CommutationAFinalResult}
\end{align}
where we have defined the quantities $L \equiv -ib$ and $R \equiv -ic$.


Starting from \eqref{eq:CommutationAFinalResult} we will soon derive a relation between the bare and renormalized masses below but, before procedding, it is important to establish a non-trivial connection between the spectral functions $\rho(s)$ and $\tilde{\rho}(s)$.  As usual, we shall decompose the spectral functions in their discrete and continuum parts
\begin{equation}
	\rho(s) = Z \delta (s - m^2_r) + \sigma(s), \quad\quad \tilde{\rho}(s) = s^{-1/2}\tilde Z \delta (s - m^2_r) + s^{-1/2}\tilde{\sigma}(s).
	\label{eq:DiscreteContinuousContributions}
\end{equation}
From equation \eqref{RelationBetweenSpectralFunctions} and its general solution \eqref{eq:SpectralRepresentation1}, it is possible to compute the vacuum expectation value for the matter current. In fact,
\begin{align}
    \langle 0 | \left[ J_{\mu}(x), J_{\nu}(y) \right] | 0 \rangle = -i\int_0^{\infty} ds \ s\left(s-m^2\right) \rho_{\mu \nu}(x,y;s),
\label{MatterCurrent}
\end{align}
where we have defined the spectral density, $\rho_{\mu \nu}(x,y;s)$, of the matter current as 
\begin{equation}
     \rho_{\mu \nu}(x,y;s) = \left[ \rho(s) \left( \eta_{\mu\nu} + s^{-1}\partial_{\mu}\partial_{\nu} \right) + \widetilde{\rho}(s) \epsilon_{\mu\nu\beta}\partial^{\beta} \right] \Delta(x-y;s).
\end{equation}
The form of \eqref{MatterCurrent} was, of course, expected by construction. Imposing the constraint (\ref{AsymptoticConditionForCurrents}), we obtain that the following relation holds asymptotically
\begin{equation}
    \epsilon^{\nu\alpha\mu} \partial_{\alpha}\langle 0 | \left[ J^{\text{as}}_{\mu}(x), J^{\text{as}}_{\nu}(y) \right] | 0 \rangle = - m_r \langle 0 | \left[ J_{\text{as}}^{\nu}(x), J^{\text{as}}_{\nu}(y) \right] | 0 \rangle.
\end{equation}
Thus, choosing only the discrete parts in \eqref{eq:DiscreteContinuousContributions} we have
\begin{equation}
    \int_0^{\infty} ds \ s\left(s+m^2\right) \left[ s^{-1/2}\tilde Z\epsilon^{\nu\alpha\mu}\epsilon_{\mu\nu\beta}\partial_{\alpha}\partial^{\beta} + 2m_r Z \right] \delta(s - m^2_r) \Delta(x-y;s) = 0,
\end{equation}
from which it follows that
\begin{equation}
    \tilde Z = \text{sgn}(m_r)Z.
\end{equation}
For completeness, after plugging this result back in equation (\ref{eq:DiscreteContinuousContributions}), we get from (\ref{spectral2}) that
\begin{equation*}
	L = \frac{Z}{m_r^2} + \int^{\infty}_{0} ds \ \frac{\sigma(s)}{s}, \quad\quad R = \frac{Z}{m_r} + \int^{\infty}_{0} ds \ s^{-1/2}\tilde{\sigma}(s), \quad\quad 1 = Z + \int^{\infty}_{0} ds \ \sigma(s).
\end{equation*}

Now, acting with the differential operator $\Box\eta^{\mu\gamma} +m\epsilon^{\mu \beta \gamma}\partial_\beta$ on the two-point function (\ref{eq:CommutationAFinalResult}) we obtain for the left-hand side, by using the equations of motion (\ref{eq:GaugeCondition}) and (\ref{eq:EquationForCurrent}), the following result\footnote{The ellipsis is the result of the unequal-time commutator between the interacting Abelian gauge field and an arbitrary matter current. Although, it is not known, we do not need the explicit result to derive equation \eqref{eq:MassRenormalization} because charged fields commute with the Abelian gauge field at equal-time.} 
\begin{align}
\langle 0 | \left[ (1-\alpha)\partial^{\gamma}B(x) - J^{\gamma}(x), A_{\nu}(y) \right] | 0 \rangle &= (1-\alpha) \partial^{\gamma}_x\langle 0 | \left[ B(x) , A_{\nu}(y) \right] | 0 \rangle - \langle 0 |\left[ J^{\gamma}(x) , A_{\nu}(y) \right] | 0 \rangle \nonumber \\
&= i(1-\alpha)\partial^{\gamma}\partial_{\nu}\Delta(x-y;0) + \cdots.
\end{align}
Thus, together with similar manipulations for the right-hand side, we have
\begin{align}
    i(1-\alpha)\partial^{\gamma}\partial_{\nu}\Delta(x-y;0) + \cdots &= -imR\partial^{\gamma}\partial_{\nu}\Delta(x-y;0) -i\alpha\partial^{\gamma}\partial_{\nu}\Delta(x-y;0) \nonumber \\
    &\quad-i\int_0^{\infty} ds\  \rho(s) \left( -s\delta^{\gamma}_{\nu} - \partial^{\gamma}\partial_{\nu} + m\epsilon_{\nu}^{~\beta\gamma}\partial_{\beta} \right)\Delta(x-y;s) \nonumber \\
    &\quad-i\int_0^{\infty} ds\  \widetilde{\rho}(s) \left( -s\epsilon^{\gamma}_{~\nu\beta}\partial^{\beta} - m\partial^{\gamma}\partial_{\nu} - ms\delta^{\gamma}_{\nu} \right)\Delta(x-y;s) \nonumber.
\label{PreResult}
\end{align}
After considering the spatial components $\gamma, \nu = i, j$ at equal times and using the Cauchy data (\ref{CauchyData}), it follows that
\begin{equation}
    0 = im\epsilon_{j}^{~0i}\delta^2(\Vec{x}-\Vec{y})\int_0^{\infty}ds\ \rho(s) - i\epsilon^{i}_{~j0}\delta^2(\Vec{x}-\Vec{y})\int_0^{\infty} ds\ s \tilde{\rho}(s) ,
\end{equation}
or
\begin{equation}
    m = \int_0^{\infty} ds\ s \tilde{\rho}(s).
\end{equation}
Replacing \eqref{eq:DiscreteContinuousContributions} in this result we get straightforwardly that
\begin{equation}
	m = Zm_r + \int^{\infty}_0 ds \ s^{1/2}\tilde{\sigma}(s). 
	\label{eq:MassRenormalization}
\end{equation}

This is the most important result of this paper. We interpret (\ref{eq:MassRenormalization}) as a \textit{non-perturbative model-dependent} relation between the bare and renormalized mass of the photon. It shows a new property which is intimately related to the dimensionality of the model. In fact, in the limit of vanishing bare mass $m\rightarrow 0$, the renormalized photon mass $m_r$ \emph{does not a priori vanish} and it depends on the continuous part of the spectral function $\tilde{\rho}(s)$ which arose only because we were working in 2 + 1 dimensions. It is worthwhile to mention that a similar equation relating the renormalized with the bare mass arises in 3 + 1 dimensions, the so-called Johnson's theorem. However, in that case, we conclude that in the limit $m\rightarrow 0$, the renormalized mass must vanish unless the matter current has massless discrete spectrum. A well-known example for the latter statement occurs in the presence of spontaneous symmetry breaking where gauge bosons can be massive \cite{Nakanishi}.


The next step is to identify what kind of matter current may produce a non-vanishing $\tilde \sigma(s)$. Certainly, it must break discrete symmetry even in the limit of vanishing bare mass since we are interested in dynamical mass generation. Although \cite{mcs} mentioned an explicit perturbative non-discrete symmetry breaking example in which scalar matter has nonvanishing $\tilde \sigma(s)$, it turns out that it is proportional to the bare mass, thus, the photon remains massless in the presence of scalars. Consequently, we are left with massive fermions in bidimensional representation. Since the source of the parity breaking comes from the mass term in the Dirac Lagrangian, it is expected that the topological mass generation depends strongly on the fermion mass. In the next section, our assumptions are verified perturbatively and in section 4 we show that we arrive at a massless discrete pole structure when considering $m_r \to 0$. In fact, the specific  low energy prescription used to manipulate the equations $(6)$ and $(11)$ loses its sense in the limit $m_r \to 0$ since we cannot postulate an asymtoptic excitation such as $ \mathcal{U}^{\mu}(x)$ that explicitly violates parity without a discrete symmetry breaking Lagrangian. It can be perturbatively shown that without topological as well as fermion bare masses they are not radiatively generated \cite{jac}. On the other hand, in the presence of any of those terms, the other is dynamically obtained. Since they break discrete symmetries, the previous discussion is in agreement with the Wilsonian perspective.

\section{Perturbation Theory}
Let us denote the following smooth limit 
\begin{equation}
	\lim_{m \to 0}\tilde \rho(s) = \tilde \rho (s)_{\text{QED}_3},
\end{equation}
where the right-hand side represents the desired QED$_3$ parity breaking contribution. We can extract from the computations made for the vacuum polarization tensor in QED$_3$ using causal perturbation theory \cite{Pim1} the following result 
\begin{equation}
	\lim\limits_{m\to 0} \int ds \ s^{1/2}\tilde{\sigma}(s) = \text{Im} \left( \frac{e^2m_e}{4\pi ^2} \int_{4m_e^2}^{\infty} ds \   s^{-3/2} \log \left( \frac{1 - \sqrt{s/4m_e^2}}{1 + \sqrt{s/4m_e^2}} \right) \right).
\end{equation}
As discussed in the previous section, the continuous part is non-vanishing in the limit of $m \to 0$ due to the presence of the electron mass $m_e$ which manifests as a symmetry breaking term. Using \eqref{eq:MassRenormalization} we obtain the one-loop result
\begin{equation}
    m_r = \frac{e^2}{4\pi}\text{sgn}(m_e).
\end{equation}

\section{Asymptotic Structure}
Having established the non-perturbative description of the phenomenon of dynamical mass generation of a gauge field through interactions with matter in 2 + 1 dimensions, we are ready to perform an analysis of the asymptotic structure of the theory.

First, we extract the discrete spectrum of (\ref{eq:CommutationAFinalResult}) assuming asymptotic completeness \cite{mcs}
\begin{align}
	\langle 0 | \left[ A_{\mu}(x), A_{\nu}(y) \right] | 0 \rangle &\xrightarrow{\text{Disc. Spectr.}} i \left( L\partial_{\mu}\partial_{\nu} - R\epsilon_{\mu\nu\beta}\partial^{\beta} \right) \Delta(x-y;0) -i\alpha\partial_{\mu}\partial_{\nu}E(x-y;0) 	\nonumber \\ 
	&\qquad\qquad\quad- i Z  \left( \eta_{\mu\nu} + \frac{1}{m_r^2}\partial_{\mu}\partial_{\nu}  - \frac{1}{m_r} \epsilon_{\mu\nu\beta}\partial^{\beta} \right) \Delta(x-y;m_r^2).
\end{align}
We next define the asymptotic field of the Heisenberg operator $A_{\mu}$ as $A_{\mu}^{\text{as}} = Z^{-1/2}A_{\mu}$ and the renormalized gauge parameter as $\alpha_r = Z^{-1} \alpha$ in terms of which the commutator for $A_{\mu}^{\text{as}}$ reads
\begin{multline}
	\big[ A_{\mu}^{\text{as}}(x), A_{\nu}^{\text{as}}(y) \big] = \\ i \left[ \left( \frac{1}{m_r^2} + Z^{-1}\int^{\infty}_{0} ds \ \frac{\sigma(s)}{s} \right) \partial_{\mu}\partial_{\nu} - \left( \frac{1}{m_r} + Z^{-1}\int^{\infty}_{0} ds \ s^{-1/2}\tilde{\sigma}(s) \right) \epsilon_{\mu\nu\beta}\partial^{\beta} \right] \Delta(x-y;0)  	 \\ 
	\quad-i\alpha_r\partial_{\mu}\partial_{\nu}E(x-y;0) - i \left( \eta_{\mu\nu} + \frac{1}{m_r^2}\partial_{\mu}\partial_{\nu}  - \frac{1}{m_r} \epsilon_{\mu\nu\beta}\partial^{\beta} \right) \Delta(x-y;m_r^2).
	\label{eq:CommutationAAsymptotic}
\end{multline}
In view of (\ref{eq:B-Field}) we define the asymptotic field $B^{\text{as}} = B$ because it is just a free field.

Having determined (\ref{eq:CommutationAAsymptotic}), we are in position to distinguish between massive and massless spectrum by decomposing $A_{\mu}^{\text{as}}$ in terms of the following fields
\begin{equation}
	\mathcal{U}^{\mu} = \frac{1}{m_r} \left( \epsilon^{\mu\nu\sigma}\partial_\nu A_\sigma^{\text{as}}- \frac{\partial^\mu B^{\text{as}}}{m_r} \right) , \qquad  \cal{A}^\mu = A^\mu_{\text{as}}- \tilde{\cal{U}}^\mu.
\end{equation}
The non-physical part $\cal{A}^\mu$ is purely massless while the transverse part is physical, massive and its commutator is given by
\begin{equation}
	\big[ \mathcal{U}_{\mu}(x),\mathcal{U}_{\nu}(y) \big] = -i \left( \eta_{\mu\nu} + \frac{1}{m_r^2}\partial_{\mu}\partial_{\nu}  - \frac{1}{m_r} \epsilon_{\mu\nu\beta}\partial^{\beta} \right) \Delta(x-y;m_r^2).
\end{equation}
Note that this expression recovers the physical Hilbert space of the MCS theory. Therefore, we conclude that the \textit{Chern-Simons mass term has been induced by the interaction of the photon with matter}. This result is compatible with the discussion given after equation (\ref{Trick}) since $\cal{U}^\mu$ represents our massive pole.






The important point of our result is that this phenomenon does not occur via an ``eating" process. In fact, it is an intrinsic characteristic of the dimensionality and the topological properties of the model. The fermionic and gauge degrees of freedom must remain the same separately. It means that the latter can not have both massive and massless poles in order to preserve its degrees of freedom before and after the interaction. It is known that in 2 + 1 dimensions both MCS and Maxwell fields have one local excitation due to its Hamiltonian similarity. We have shown that the massive excitation is physical in the sense of $ \big[ {\cal{U}}^\mu(x), B(y) \big]=0$. So it must represent the unique observable degree of freedom.

It is also important to mention that the emergence of a Chern-Simons term can be understood as a topological Higgs mechanism \cite{top}. It is expected since every mass generation can be expressed as a kind of Higgs phenomenon \cite{Nakanishi}.

Furthermore, we can show that in the massless limit the asymtoptic field recovers the well-known discrete massless pole structure. To see this, we use the definition of the renormalized mass and its Taylor expansion given by
\begin{equation}
	\Delta(x-y,m_r)=\Delta(x-y,0)-E(x-y,0)m_r^2+ \cdots.
\end{equation}
After the redefinition of variables
\begin{equation}
	A_{\mu}^{\text{as}}(x)\to A_{\mu}^{\text{as}}(x)-\frac{1}{2}\left( Z^{-1}\int^{\infty}_{0} ds \ \frac{\sigma(s)}{s} \right)\partial_{\mu}B^{\text{as}}(x),
\end{equation}
we get \cite{Nakanishi}
\begin{align}
	\left[ A_{\mu}^{\text{as}}(x), A_{\nu}^{\text{as}}(y) \right] =   	
	-i\alpha_r\partial_{\mu}\partial_{\nu}E(x-y;0) - i \left( \eta_{\mu\nu}\Delta(x-y;0)-\partial_{\mu}\partial_{\nu}E(x-y;0)  \right). 
\end{align}

\section{Conclusion}
Throughout this work a dynamical mass generation for QED$_3$ was verified first by means of the Heisenberg equations of motion valid in all Hilbert space. Later, we obtained this same result by studying the asymptotic two-point structure of the renormalized photon fields whose physical part is the same as that of the Maxwell-Chern-Simons theory. This last observation allows us to talk about a dynamically generated topological mass term.

This result was previously obtained in the perturbative approach but here we had the opportunity to make some general observations which are characteristic of the non-perturbative treatment. The appearance of this massive excitation was expected because the Wilsonian perspective strongly indicates it since the addition of a Chern-Simons topological mass term is a natural generalization to QED in $D=2+1$ dimensions if we are in a parity breaking scenario. So, we also pointed out the importance of coupling with bidimensional massive fermions for the occurence of the mass generation phenomena.

The asymptotic structure was obtained and the massive excitation recovered is the one previously found by means of the operator equations of motion. We also show how to circumvent the Johnson's theorem in order to have a dynamically generated renormalized mass to the photon field. The method employed is indeed consistent since the massless structure could be continuosly reached in the limit $m_r \to 0$. 

Finally, we have pointed out throughout the introduction of this work that these models have interesting properties when studied in their dual language. It would be interesting to know how the notion of duality can be formulated within the KON formalism. This investigation is reserved to another paper \cite{future}.

\section*{Acknowledgments}
The authors would like to thank the referee for the comments and suggestions to improve the manuscript significantly. G. B. de Gracia and L. Rabanal thank CAPES for support, and B. M. Pimentel thanks CNPq for partial support.

\appendix

\section{Remarks on covariant quantization of the interacting Abelian gauge theory}
In this appendix we shall derive equation \eqref{RelationBetweenSpectralFunctions}. In section 2 we learned that in an abelian gauge theory with linear covariant gauge fixing and arbitrary matter current, $B$ satisfies a massless free-field equation \eqref{eq:B-Field}. Consequently, we can obtain an integral representation for $B(y)$
\begin{equation}
    B(y) = \int d^3z \left[ \partial_0^z \Delta(y-z;0)B(z) - \Delta(y-z;0)\partial_0 B(z) \right].
\label{BIntegral}
\end{equation}
Owing to the $z^0$ independence of \eqref{BIntegral}, we can compute four-dimensional commutation relations of the form $\left[ \Phi(x),B(y) \right]$ by using \eqref{BIntegral} evaluated at $z^0 = x^0$ and the equal-time commutations relations. In particular, we have $\left[B(x), B(y)\right] = 0$ and
\begin{align}
    \left[ A_{\mu}(x),B(y) \right] &= \left[ A_{\mu}(x), \int d^3z \left[ \partial_0^z \Delta(y-z;0)B(z) - \Delta(y-z;0)\partial_0 B(z) \right]\right] \nonumber \\
    &= -i\partial_{\mu}\Delta(x-y;0).
\label{A2}
\end{align}
for the Abelian gauge field. This suggests a remarkable similarity between the field $B(x)$ and the generator of local gauge transformations. In fact, $\left[B(x), \phi(y)\right] = \phi(x)\Delta(x-y,0)$ and $\left[\psi(x), B(y) \right] = e\psi(x)\Delta(x-y,0)$ for scalar and fermion fields, respectively. See \cite{Nakanishi} for more details. 

From \eqref{A2} it follows immediately, by the symmetry of the product of two derivatives, that
\begin{equation}
    \left[ F_{\mu\nu}(x),B(y) \right] = 0.
\label{A3}
\end{equation}
Moreover, from \eqref{eq:EquationForCurrent}, \eqref{A2} and \eqref{A3} we get
\begin{align}
    \left[ J^{\nu}(x), B(y)\right] &= -\left[ \partial_{\mu} F^{\mu\nu}(x) + m\epsilon^{\nu\mu\beta}\partial_{\mu}A_{\beta}(x) - \partial^{\nu} B(x), B(y)\right] \nonumber \\
    &= -m\epsilon^{\nu\mu\beta}\partial_{\mu}^x \left[A_{\beta}(x), B(y)\right] \nonumber \\
    &= im\epsilon^{\nu\mu\beta}\partial_{\mu}\partial_{\beta}\Delta(x-y;0) \nonumber \\
    &= 0.
\end{align}
We interpret this result as the statement of gauge invariance for the matter current. In fact, any field $\Psi(x)$ that satisfies $\left[\Psi(x), B(y)\right] = 0$ is a gauge invariant or physical field.

Now, we can proceed with the derivation of \eqref{RelationBetweenSpectralFunctions}. We start by writing the equation of motion \eqref{eq:EquationForCurrent} as follows
\begin{equation}
    \Box A^{\nu} = - J^{\nu} + (1-\alpha)\partial^{\nu}B - m\epsilon^{\nu\mu\beta}\partial_{\mu}A_{\beta}.
\end{equation}
From this it follows, by straightforward computation, that
\begin{align}
    \left[ J^{\alpha}(x), J^{\beta}(y) \right] &= \big[-\Box^x A^{\alpha}(x) + (1-\alpha)\partial_x^{\alpha}B(x) - m\epsilon^{\alpha\mu\nu}\partial^x_{\mu}A_{\nu}(x), -\Box^y A^{\beta}(y) \nonumber \\ 
    &\quad+ (1-\alpha)\partial_y^{\beta}B(y) - m\epsilon^{\beta\rho\sigma}\partial^ y_{\rho}A_{\sigma}(y) \big] \nonumber \\
    &= \left[ \Box^x A^{\alpha}(x), \Box^y A^{\beta}(y) \right] \nonumber \\
    &\quad- (1-\alpha)\Box^x\partial_y^{\beta} \big[A^{\alpha}(x), B(y) \big] - (1-\alpha)\Box^y\partial_x^{\alpha} \left[B(x), A^{\beta}(y) \right] \nonumber \\
    &\quad-  m(1-\alpha)\epsilon^{\beta\mu\nu}\partial_x^{\alpha}\partial_{\mu}^y \left[B(x),A_{\nu}(y) \right] -  m(1-\alpha)\epsilon^{\alpha\mu\nu}\partial_y^{\beta}\partial_{\mu}^x \left[A_{\nu}(y),B(x) \right] \nonumber \\
    &\quad+ m\epsilon^{\beta\mu\nu}\Box^x\partial^y_{\mu} \left[A^{\alpha}(x),A_{\nu}(y) \right] + m\epsilon^{\alpha\mu\nu}\Box^y\partial^x_{\mu} \left[A_{\nu}(x),A^{\beta}(y) \right] \nonumber \\
    &\quad+ m^2\epsilon^{\alpha\mu\nu}\epsilon^{\beta\rho\sigma}\partial_{\mu}^x\partial^y_{\rho} \left[ A_{\nu}(x), A_{\sigma}(y) \right].
\end{align}
After using \eqref{A2} together with $\Box\Delta(x-y;0) = 0$ and the fact that $\epsilon^{\alpha\mu\nu}\partial_{\mu}\partial_{\nu}\Psi$ vanishes for any appropriate function $\Psi$, the second and third line vanishes. Thus, we obtain 
\begin{align}
    \left[ J^{\alpha}(x), J^{\beta}(y) \right] &= \Box^x\Box^y \left[  A^{\alpha}(x),  A^{\beta}(y) \right] + m\epsilon^{\beta\mu\nu}\Box^x\partial^y_{\mu} \left[A^{\alpha}(x),A_{\nu}(y) \right] \nonumber \\
    &\quad+ m\epsilon^{\alpha\mu\nu}\Box^y\partial^x_{\mu} \left[A_{\nu}(x),A^{\beta}(y) \right] 
    + m^2\epsilon^{\alpha\mu\nu}\epsilon^{\beta\rho\sigma}\partial_{\mu}^x\partial^y_{\rho} \left[ A^{\nu}(x), A_{\sigma}(y) \right],
\end{align}
or more precisely,
\begin{equation}
\bigg( \Box^x\eta^{\alpha\nu} + m\epsilon^{\alpha\mu\nu}\partial_{\mu}^x \bigg) \bigg( \Box^y\eta^{\beta\sigma} + m\epsilon^{\beta\mu\sigma}\partial_{\mu}^y \bigg) \big[ A_{\nu}(x), A_{\sigma}(y) \big] = \left[ J^{\alpha}(x), J^{\beta}(y) \right].
\end{equation}

\end{document}